\documentclass[a4paper,color]{article}

\usepackage[utf8]{inputenc}   
\usepackage[T1]{fontenc}      
\usepackage{geometry}         
\usepackage[english]{babel} 
\usepackage{subfig}
\usepackage{caption}
\usepackage{graphicx} 
\usepackage{epstopdf} 
\usepackage{amsmath} 
\usepackage{color}
\usepackage{float}          
\usepackage{url}          

\title{A morphological study of cluster dynamics between critical points}           

\author{Thibault Blanchard, Leticia F. Cugliandolo and Marco Picco\\
Laboratoire de Physique Th\'eorique et Hautes Energies, \\
Universit\'e Pierre et Marie Curie - Paris VI, \\
4 Place Jussieu, 75252 Paris Cedex 05, France
}

\date{\today}

\relax

\def\beq{\begin{equation}}
\def\eeq{\end{equation}} 
\def\bea{\begin{eqnarray}} 
\def\eea{\end{eqnarray}} 
\relax

\begin{document}

\maketitle

\begin{abstract}   
We study the geometric properties of a system initially in equilibrium at a critical point that is suddenly quenched to
another critical point and subsequently evolves towards the new equilibrium state. We focus on the bidimensional Ising
model and we use numerical methods to characterize the morphological and statistical properties of spin  and
Fortuin-Kasteleyn clusters during the  critical evolution. The analysis of the dynamics of an out of equilibrium
interface is also performed. We show that the small scale properties, smaller than the target critical growing length
$\xi(t) \simeq t^{1/z}$ with $z$ the dynamic exponent, are characterized by equilibrium at the working critical point,
while the large scale properties, larger than the critical growing length, are those of the initial critical point.
These features are similar to what was found for sub-critical quenches. We argue that quenches between critical points
could be amenable to a more detailed analytical description.
\end{abstract}

\newpage

\tableofcontents            

\newpage

\section{Introduction}

The relaxation dynamics of a macroscopic system taken to its critical point with some quenching protocol has received
much attention. Most of the existing studies focused on the time evolution of global quantities (magnetization,
susceptibility, correlation functions, etc.) and characterized their scaling properties with numerical simulations and
renormalization group techniques~\cite{hohenberg_theory_1977,Jan92,calabrese_ageing_2005,tauber_critical_2012}. These
studies are not restricted to any spatial nor order parameter dimensionality.

In {\it two dimensional} critical systems in equilibrium, powerful theoretical tools such as Coulomb gas
methods~\cite{nienhuis_coulomb_1987}, conformal field theory (CFT)~\cite{cardy_conformal_1987} and stochastic Loewner
evolution (SLE)~\cite{Cardy05,gruzberg_stochastic_2006,Bernard06}, allowed one to characterize the geometric and
statistical properties of a large variety of mesoscopic observables in great detail. These objects give a more complete
image of the system's equilibrium configurations than the global observables accessed with scaling arguments and
renormalization group techniques.  However, as far as we know, nothing is known about these objects during the out of
equilibrium evolution of the same (and other) critical samples.

An extensive numerical and analytic investigation of  the {\it coarsening sub-critical dynamics} of two dimensional
models from a mesoscopic point of view was carried out in recent years.  The models treated were the  clean Ising model
with non-conserved~\cite{Arenzon2007,Sicilia2007,Barros09} and conserved~\cite{Sicilia2009} order parameter dynamics,
the random ferromagnet with non-conserved order parameter dynamics~\cite{Sicilia2008} or still the $q$ state Potts
model~\cite{loureiro_curvature_2010,loureiro_geometrical_2012}. These studies allowed one to build a rather complete
picture of the geometric and statistical properties of the spin clusters in these bidimensional systems. More precisely,
their domain and hull-enclosed areas as well as their boundary lengths and the relation between areas and perimeters
were analyzed and characterized in detail.

The aim of this work it to present a similar study of relevant dynamic geometric objects during the critical
non-equilibrium evolution of the $2d$ Ising model evolving from equilibrium at another critical point, in this case an
infinite temperature configuration that is equivalent to critical uncorrelated site percolation. We use simple scaling
arguments and extensive numerical simulations.

Let us be more specific about the objects of our study.  The most natural objects to consider are the spin clusters,
i.e. clusters of nearest neighbor spins on the lattice that point in the same direction.  These clusters are accessible
via direct observation of the system. However, at the critical point, they are not appropriate to describe the critical
equilibrium properties of the Ising model since their shape and statistical properties are not only dictated by the
physical correlations but also by purely geometrical factors~\cite{sykes_note_1976,coniglio_clusters_1980}.  To remedy
this problem one has  to consider, in place,  smaller clusters, namely the Fortuin-Kasteleyn (FK) ones, that capture
exclusively the equilibrium  physical correlations in the system.  Although not relevant to describe the equilibrium
physical macroscopic properties of the samples the spin clusters are, nonetheless, also critical at the phase
transition. Consequently, they are characterized by a different set of exponents from the ones of the FK clusters. For
this reason (to be  explained more thoroughly in Sec.~\ref{conf}) spin clusters are also interesting to analyze and we
study both types of clusters here.

The article is organized as follows. In Sec.~\ref{sec:definitions} we list the definitions of the different objects we
study. The concrete analysis is presented next.  We first check that the dynamical exponent $z$ governing the critical
dynamics of spin clusters is actually the one of the $2d$ Ising model, as obtained with other means, as the dynamic
renormalization group method~\cite{Jan92}. To the best of our knowledge, this has never been brought to light before.
We then study in full extent the number densities of various quantities giving access to the structure of spin clusters
on a large interval of sizes. The validity of the dynamical scaling hypothesis is tested upon these number densities.
Relations between areas and boundaries are also explored. The same method is applied to the FK clusters. We later turn
our attention to the dynamics of a single interface and the comparison with the equilibrium results given by conformal
field theory and stochastic Loewner evolution. This is the content of Sec.~\ref{sec:dynamic-scaling}. Finally, in
Sec.~\ref{sec:conclusions} we draw our conclusions and we discuss some lines of future research.

\section{Definitions}
\label{sec:definitions}

\subsection{Spin and Fortuin-Kasteleyn clusters}

Simple scaling arguments from percolation theory~\cite{stauffer_introduction_1994} suggest that, in equilibrium, the
divergence of the mean size of some kind of finite cluster at the critical point should be governed by the
susceptibility exponent or, equivalently, the one of the spin-spin correlation function, that is the same as the
probability for two spins to be in the same such cluster.  The ensemble of nearest-neighbor spins that are parallel to
each other constitute a {\it domain} or {\it spin cluster}. These are the most natural geometric objects, the critical
properties of which one would expect to be linked to the ones of macroscopic physical observables. However, spin
clusters  do not capture the underlying critical properties of statistical physics models. This fact was first noticed
by Sykes and Gaunt~\cite{sykes_note_1976} who remarked that the mean size of the finite domains of the Ising model on a
two dimensional lattice (2dIM) diverges at the critical point with an exponent that is different from the magnetic
susceptibility exponent (see Table~\ref{tab_exp}).  Moreover, at infinite temperature there are spin clusters of
arbitrary large size although the spin-spin correlation function vanishes. Still, for the particular case of the 2dIM,
domains are critical at the critical point although they do not give access to the relevant critical exponents of the
model.  We will come back to this point in Sec.~\ref{conf}. In three dimensions, this is not the case and these spin
clusters do not even percolate at the critical point~\cite{mueller-krumbhaar_percolation_1974}.

\begin{figure}[ht!] 
       	\begin{center} 
        	\includegraphics[scale=1]{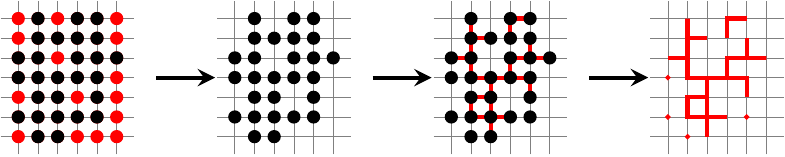}
        	\end{center} 
        	\caption{Sketch of the construction of a FK cluster. A domain is identified. The bonds between nearest-neighbor 
		aligned spins (represented with black dots on the lattice sites) are erased using the FK procedure.
		Two FK disconnected clusters remain. The surviving bonds are highlighted on the edges of the lattice.
	\label{def:fk}}
\end{figure}

The point was then to build clusters containing just physical correlations between spins (and not trivial geometric
ones).  This is achieved by the \emph{Fortuin-Kasteleyn clusters}~\cite{fortuin_random-cluster_1972}, proposed by these
authors as a way to describe with the {\it random cluster model} percolation, the Ising model, the Potts model and many
other statistical models. Independently, Coniglio and Klein~\cite{coniglio_clusters_1980} linked the critical properties
of the Ising model to the percolation of \emph{Ising droplets} which were later identified with the Fortuin-Kasteleyn
clusters, the name that prevailed in the literature. The Fortuin-Kasteleyn clusters are constructed as follows. Starting
with a spin domain, one first draws all bonds linking nearest-neighbor spin on the cluster and then erases bonds with a
temperature dependent probability $e^{-2K}$. In such a way, the original bond-cluster typically diminishes in size and
may even get disconnected.  More precisely, the construction of these clusters from the partition function ${\cal Z}$ is
the following.  One rewrites the exponential of the sum as a product of exponentials: 
\begin{equation} 
{\cal Z}=\sum_{\{\sigma\}}e^{K\sum_{\langle i,j\rangle} \sigma_i\sigma_j}=\sum_{\{\sigma\}}\prod_{\langle i , j \rangle}e^{K\sigma_i\sigma_j}, 
\end{equation} 
with $K=J/(k_B T)=\beta J$, $J$ the exchange coupling and $k_B$ the Boltzmann constant. Hereafter we measure temperature
$T$ in units of $J/k_B$.  By remarking that, since $\sigma_i\in\{-1;1\}$, 
\begin{equation} 
e^{K\sigma_i\sigma_j}=e^K \ [(1-e^{-2K})\delta_{\sigma_i,\sigma_j}+e^{-2K}],
\end{equation} 
${\cal Z}$ can be recast in the form: 
\begin{equation}
{\cal Z}=e^{mK}\sum_{\{\sigma\}}\prod_{\langle i , j \rangle} [p\delta_{\sigma_i,\sigma_j}+(1-p)] ,
\end{equation}
with $m$ the total number of bonds of the lattice ($m=3L^2$ for a planar triangular lattice of linear size $L$) and
$p=1-e^{-2K}$. By expanding the product, if $\sigma_i=\sigma_j$ one keeps the bond  $\langle i , j \rangle$ with
probability $p$ or erases it with probability $1-p$. The broken bonds ($\sigma_i = -\sigma_j$) are never kept. A
connected collection of kept bonds is a Fortuin-Kasteleyn (FK) cluster. This whole construction is called the
Fortuin-Kasteleyn random cluster model and an example is presented in Fig.~\ref{def:fk}.  These clusters possess the
searched properties.  Their exponents are those of the Ising model~\cite{jan_study_1982}. For example, the probability
for two spins to belong to the same FK cluster is exactly the spin-spin correlation of the Ising model.  Note that the
FK construction can be easily extended to all $q$-state Potts models, even with non integer $q$, in any dimensions.

\subsection{A few hints on conformal field theory}
 \label{conf}
The existence of two types of critical clusters, the spin and FK ones belonging to distinct universality classes is
valid for all $q$-state Potts model with $0\le q\le 4$ in two dimensions.  In order to understand this point it is
useful to consider the Coulomb gas formulation of the Potts model~\cite{nienhuis_coulomb_1987}. In this formulation, the $2d$
$q$-state Potts model can be described by the  parameter $\kappa\in[4,8]$ such that
\begin{equation}
\sqrt{q}=-2\cos\left(\frac{4\pi}{\kappa}\right).
\label{eq_coulomb}
\end{equation}
We choose a convention such that $\kappa$ also corresponds to the SLE parametrization for the interfaces associated to
the $q$-state Potts model. We will come back to this point in Sec.~\ref{sec:interface} where we will study the behaviour
of an out of equilibrium interface. The parameter $\kappa$ is also related to the central charge of the corresponding
conformal field theory by the relation~\cite{nienhuis_coulomb_1987,cardy_conformal_1987}: 
\begin{equation}
\mathrm{c}=\frac{1}{4}(6-\kappa)\left(6-\frac{16}{\kappa}\right).
\label{eq:central-charge}
\end{equation}

For $\kappa \in [8/3,4]$ this Coulomb gas representation describes another class of models,  the {\it tricritical Potts
model}, i.e. Potts models with dilution. Note that the central charge is invariant under the transformation $\kappa
\rightarrow 16/\kappa$ which maps a critical $q$-state Potts model onto a tricritical Potts model with a different
number of states given by eq.~(\ref{eq_coulomb}). This means that there exist two critical theories for a given central
charge. One is associated to the critical Potts model and the relevant structures are the FK clusters. The other one is
the tricritical Potts model and the relevant clusters are the domains or spin clusters. This duality is such that the
geometrical clusters of one model are the FK clusters of the other model and vice
versa~\cite{duplantier_conformally_2000,janke_geometrical_2004}.

The tricritical model associated to the $2d$ Ising model ($\kappa=16/3$) is the dilute  $q=1$ Potts ($\kappa=3$).  Both
models possess the same central charge $\mathrm{c}=1/2$ but the same quantities in the two model are not associated to
the same operators of the CFT.  In $2d$ the spin clusters of the Ising model are the critical objects of the dilute
$q=1$ Potts model~\cite{stella_scaling_1989}. This explains why the percolation exponents associated to the spin
clusters are not related to the Ising model exponents but in fact are those of the tricritical $q=1$ Potts model,
described by $\kappa=3$.

The critical exponents can then be expressed in term of $\kappa$. For instance, the fractal dimensions $D_c$, $D_h$
and $D_{ep}$ of the cluster area, its hull and its external perimeter (a precise definition of these quantities will
be given in Sec.~\ref{subsec:definitions}) read~\cite{duplantier_exact_1989,duplantier_conformally_2000}:
\begin{eqnarray}
D_c &=& 1+\frac{3\kappa}{32}+\frac{2}{\kappa}
,
\\
D_{h} &=& 1+\frac{\kappa}{8} 
,
\label{sle:dim}
\\
D_{ep} &=&1+\frac{2}{\kappa}
. 
\end{eqnarray}
Note that the duality $\kappa\rightarrow 16/\kappa$ relates the fractal dimension of the external perimeter to the
dimension of the hull of the dual model, such as $D_{ep}(\kappa)=D_{h}(16/\kappa)$ for $\kappa \ge 4$.  This relation is
called Duplantier duality~\cite{duplantier_conformally_2000}. For $\kappa < 4$ the external perimeter coincides with the
hull since they are no fjords. In this case the relation between $D_{ep}$ and $D_{h}$ mentioned above does not hold
anymore and in fact $D_{ep}=D_{h}$.
\vspace{0.25cm}

\begin{table}[h]
	\centering
	{\renewcommand{\arraystretch}{1.5}
	\renewcommand{\tabcolsep}{0.2cm}
	\begin{tabular}{c|cccccc}
		\hline
		\hline
		                        & $\mathrm{c}$&$\kappa$& $q$ & $D_c$ & $D_h$& $D_{ep}$ \\
		\hline 
		percolation             &  0   & 6    & 1  &  91/48 & 7/4  & 4/3   \\
		Ising FK clusters       &  1/2 & 16/3 & 2  &  15/8  & 5/3  & 11/8  \\ 
		Ising spin clusters     &  1/2 & 3    & 1  & 187/96 & 11/8 & 11/8   \\ 
		\hline
		\hline
	\end{tabular} }
	\caption{Central charge, $\mathrm{c}$, Coulomb gas or SLE parameter, $\kappa$, Potts model parameter, $q$ [all 
		 these related by eqs.~(\ref{eq_coulomb}) and~(\ref{eq:central-charge})], and three fractal  
	         dimensions at the percolation threshold, and for FK and spin clusters at the critical point of the $2d$ 
		 Ising model. The \emph{c} subscript in the fractal dimension is for the cluster mass, the \emph{h} one for 
		 its hull, and \emph{ep} for its external perimeter.
	\label{tab_exp}
	}
\end{table}

\vspace{0.25cm}

Finally, let us note that the existence of two types of critical clusters at the critical point is specific to the $2d$
Potts model. While the FK clusters can be defined for more general models, nothing ensures that they are critical at the
transition point. For instance,  FK clusters are not critical at the critical point in $2d$ parafermionic
models~\cite{picco_geometrical_2009}. Concerning the spin clusters, it is not even sure that they are critical at the
critical point as in the $3d$ Ising model we have already mentioned.

\subsection{Definitions of the quantities computed}
\label{subsec:definitions}

\begin{figure}[ht!] 
        	\begin{center} 
        	\includegraphics[scale=1]{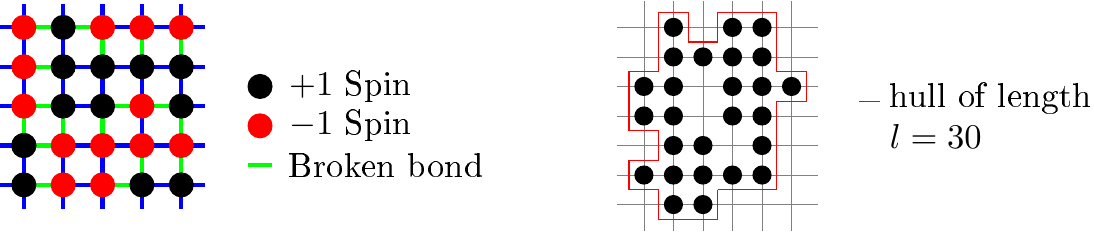}
        	\end{center} 
        	\caption{Left: sketch of an Ising spin configuration with the two-valued spins represented by black 
			 and red (black and grey) dots on the square lattice sites. Broken bonds are drawn with green 
			 (light grey) lines on the edges of the lattice. Right: a domain is singled out and its hull is 
			 represented with a thin red line. \label{def:bonds_hulls}}
\end{figure}

In this section we define the quantities that we will consider in this work. We call a \emph{domain} or \emph{spin
cluster} a connected set of sites with spins taking the same value.  The \emph{mass} (area in $2d$) of a domain is the
number of sites belonging to it.  A \emph{broken bond} is a link of the lattice between two neighboring spins with
different value (see Fig.~\ref{def:bonds_hulls}).\footnote{In all the figures we use a square lattice for simplicity.
The extension to the triangular lattice considered later should be straightforward.}
 
The \emph{domain wall} of a spin cluster is its external and internal contour, constructed as follows. One first 
generates a dual lattice by placing a site at the center of each plaquette of the original lattice. Next, the links on
the dual lattice that cross broken bonds on the original lattice are joined together. In this way, one finds a closed
loop on the dual lattice that runs along the internal or external boundary of a spin cluster in the way sketched in
Fig.~\ref{def:bonds_hulls} for the external component. There is not much theoretical knowledge on this object and it is
therefore more convenient to study other geometrical quantities which have received a theoretical description in
equilibrium.  

The \emph{hull} of a cluster is restricted to the external part of the contour, that is to say, one excludes the
contribution of the holes of the cluster (see Fig.~\ref{def:bonds_hulls}). The \emph{hull enclosed area} is the area,
i.e. the total number of sites, inside the hull (the holes within the domains are thus filled). In the example in the
right of Fig.~\ref{def:bonds_hulls} the hull-enclosed area is 29.

The length of the different types of contours defined above and living on the dual lattice is computed by counting the
number of broken bonds corresponding to the object of concern that are crossed by the boundary. We can also define the
\emph{external perimeter} built by closing the narrow gates of the hull, making in this way a smoother version of the
contour by eliminating the deep fjords. The meaning of narrow will be discussed later. As an example, on a square
lattice, a cluster composed of a unique spin has a hull of length $4$ and an area equal to $1$ while for a two-spin
cluster the hull length is $6$ and the area equals 2.

\begin{figure}[ht!] 
        \begin{center} 
        \includegraphics[scale=1]{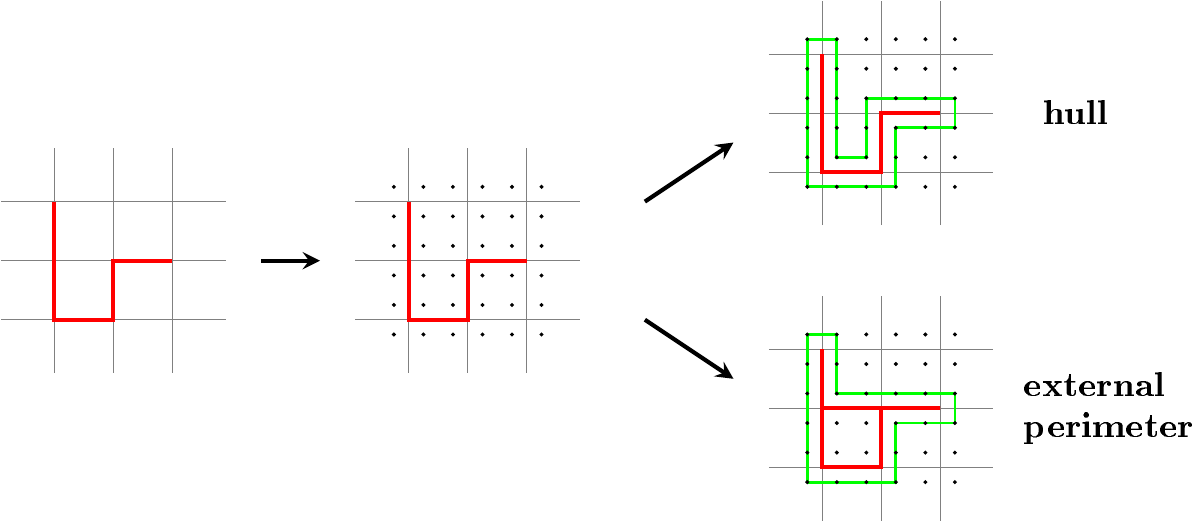}
        \end{center} 
        \caption{Left: an example of an FK cluster. Center: the sites of the sub-lattice associated to the 
	original square lattice are added as small points. Right: the hull and the external perimeter of the 
	chosen FK cluster are shown with a thin green line. 
\label{fk:hull}}
\end{figure}

The FK clusters are defined on the edges of the lattice and not on the sites of the lattice as for the spin clusters.
Therefore the length of their contour on the dual lattice is not proportional to the number of broken bonds. Indeed some
bonds linking two sites of a FK cluster may not be within the cluster if they have not been kept in the construction of
the cluster. In order to define the border of an FK cluster we then use a different sub-lattice, with four sites
associated to an original one as shown in Fig.~\ref{fk:hull}. The distance between two nearest-neighbor sub-sites counts
as the unit of length for the contour. As an example, the cluster in Fig.~\ref{fk:hull} has a hull length of 24.

In Fig.~\ref{fk:hull} we also present the construction of the external perimeter of an FK cluster. To measure its length, 
the procedure is the following: all the bonds between nearest-neighbor sites (on the original lattice)
belonging to the FK cluster are drawn, and then the walker is allowed to wander around this new cluster, that is to say, on the 
sites of the sub-lattice introduced before. The external perimeter is a smoother version of the hull and it has, consequently, 
a smaller fractal dimension. The cluster in Fig.~\ref{fk:hull} has an external perimeter of length 20 which is 
effectively smaller than its hull length.

\subsection{Statistical and geometric properties in equilibrium}	
To describe the statistical and geometrical properties of the spin and FK clusters we used the tools of percolation
theory (see e.g.~\cite{stauffer_introduction_1994}). For example, we counted the number of spin clusters with a given
area for numerous samples and we thus obtained, after normalization, the number density of the spin cluster areas. We
followed this procedure for the hull lengths, the hull-enclosed areas, etc. In general, when we consider the probability
distribution per lattice site $n_x(X)$ (of a given geometrical object $x$ taking values $X$) which follows a power law,
we denote the related critical exponent $\tau_X$, i.e.,
\begin{equation}
n_x(X)\sim X^{-\tau_X}.
\end{equation}
Either the clusters or their boundaries are fractal objects at the critical point. Their Hausdorff dimension (referred
hereafter as \emph{fractal dimension}) is non-trivial and has been related to other critical exponents. Indeed, the
fractal dimension $D_X$ of the objects described by the quantity $X$ can be expressed in terms of $\tau_X$:
\begin{equation}
D_X=\frac{d}{\tau_X-1},
\end{equation}
with $d$ the dimensionality of the lattice~\cite{stauffer_introduction_1994}. A few examples are given in Table~\ref{tab_exp}. 

In finite-size lattices  a cluster is said to percolate whenever it spans over a distance that is larger than the linear
size $L$ of the system, in at least one direction of the lattice.  The hull-enclosed areas do not exhibit a fractal
structure since they have no holes. Their fractal dimension is therefore $2$. Their boundaries behave differently and
they are fractal. Furthermore, Cardy and Ziff~\cite{cardy_exact_2002} showed that the number density of hull-enclosed
areas behaves as:
\begin{equation}
n_h(A)\sim \frac{C}{A^2}
\end{equation}
for $L^2\gg A^2 \gg a^2$ with $C=1/(8\sqrt{3}\pi)$ at the percolation threshold and $C=1/(16\sqrt{3}\pi)$ for Ising spin
clusters. The microscopic cut-off $a$ is the lattice spacing.

\section{Critical dynamics}
\label{sec:dynamic-scaling}
The stochastic relaxation dynamics after a quench to the critical point has been performed with a time dependent extension
of the renormalization group method. Scaling laws for averaged global observable such as correlation functions and
others have been obtained in many different systems. The growth of an equilibration length, $\xi(t)~\sim~t^{1/z}$ with
$z$ the dynamic critical exponent, was evidenced. In this section we study the critical dynamics from a mesoscopic point
of view and we show that this growing length also plays an important role in the characterization of the statistical
properties of fluctuating quantities.

\subsection{Description of the protocol}
\label{subsec:protocol}

In this section we study the evolution of a system after a quench from $T=\infty$ to $T_c$.  We first consider spin
clusters on a triangular lattice for which $T_c=4/\ln 3$. The choice of a triangular lattice is motivated by the fact
that the infinite temperature point exactly corresponds to a site percolation critical point on this lattice. Indeed,
the site percolation threshold for this lattice is $p_c=1/2$~\cite{sykes_exact_1963} so that, if we consider the number
density of domains, the system is exactly at the percolation threshold at infinite temperature where the spins take  the
values $1$ or $-1$ with equal probability. For the quench considered, the system is initially in equilibrium at a
critical state and after the quench it evolves towards equilibrium at another critical state, the one at $T_c$.  

At infinite temperature FK clusters are not critical, actually they are trivial, for any lattice so it is immaterial to
use the triangular or any other one. For this reason, in our study we used a square lattice. For the square lattice
$T_c=2/\ln(1+\sqrt{2})$ and the infinite temperature point is \emph{not} even a critical one for the spin clusters since
here $p_c>1/2$.  (Having said this, the studies in~\cite{Arenzon2007,Sicilia2007,Barros09} showed that, somehow
surprisingly, the sub-critical Monte Carlo dynamics of such a non-critical initial state gets, after a few time steps,
very close to the critical percolation point as far as the properties of spin clusters are concerned. For instance,
their number densities rapidly develop the critical percolation tails and only later the dynamics evolve towards the
target equilibrium state at low temperature.)
 
The simulations were carried out on a lattice of linear size $L$ with periodic boundary conditions in both directions.
For the initial condition at infinite temperature, the spins were chosen randomly with equal probability of being up or
down.  Once the system was prepared in the desired initial condition, single spin updates were performed at the
temperature of the quench. A Monte Carlo time step (MCs) corresponds to $L^2$ single spin updates. The updates were
accepted or rejected via the standard Monte Carlo Metropolis scheme. We gathered around $6 \ 10^5$ independent samples.
Unless stated otherwise, the lattice used has a linear size $L=1000$ for the analysis of spin clusters and $L=320$ for
the study of FK clusters. For each sample we computed the desired quantities every $2^n$ MCs with $n\in[1,12]$. 

For the spin clusters, the algorithm distinguishes first between the internal and external part (hull) of the perimeter
and then calculates the length of the hulls, i.e. the number of broken external bonds. To do so the lattice is scanned and when a
broken bond that has not been counted yet is found, the algorithm follows the boundary in a precise direction and keeps
track (with a cumulative angle) of the path followed.  Then, depending on the sign of the angle the boundary drawn is an
external or an internal one. Some clusters (spanning ones) have a boundary with a vanishing angle.  Those
boundaries run across the system from one side to another so they are not homotopic to a point on the torus. We
checked that these clusters are sufficiently rare so that we can discard them without affecting the statistics.  For the
FK clusters the process is the same on the sub-lattice evoked above. The algorithm walks on the sub-lattice around the
FK clusters and once a contour is formed, the angle is measured to discriminate between
internal and external contours and their lengths are computed.

The number densities obtained via this Monte Carlo method present a great dispersion, especially for large clusters
which are much rarer than the small ones. This dispersion has been greatly suppressed by choosing appropriate bin sizes
to construct the histograms, and to extract from them the distributions.  

\subsection{Spin clusters}
\label{sec:spin_clusters}
We first briefly present the simulations performed to check whether the dynamical exponent governing the dynamics of
the spin clusters conforms to the one extracted from the analysis of the correlation functions that is well documented
in the literature~\cite{bausch_renormalized_1976,nightingale_monte_2000,calabrese_ageing_2005}.  
Then we present the results for the spins clusters on a triangular lattice after the quench.

\subsubsection{The dynamic exponent}

\begin{figure}[h!,scale=0.5] 
	\captionsetup[subfloat]{labelformat=empty}
	\begin{center} 
		\subfloat[]{\includegraphics[scale=1.09]{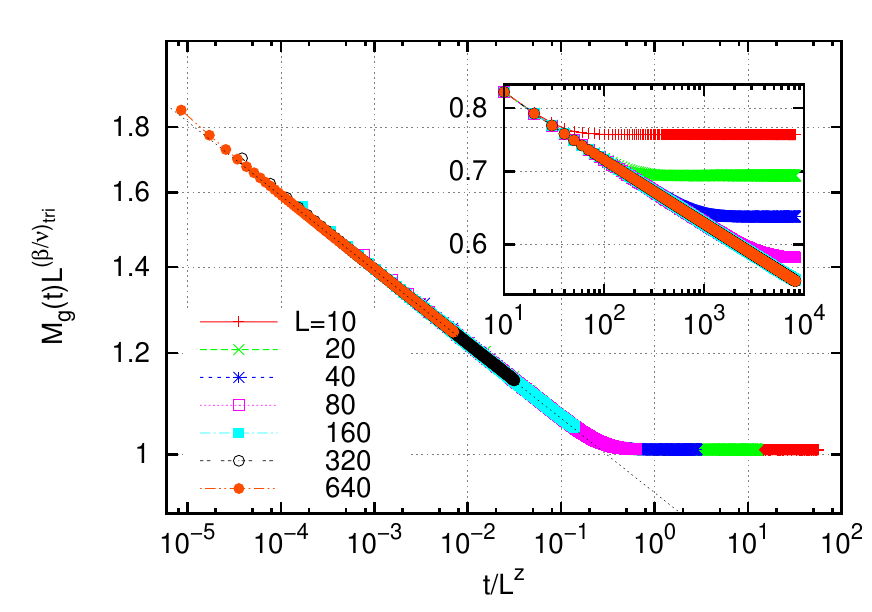}}
	\end{center} 
	\caption{Geometrical cluster magnetization $M_g(t)$. The main part shows the collapse onto a master curve with 
		 the rescaling $t\to t/L^z$ and $M_g(t)\to M_g(t)L^{(\beta/\nu)_{tri}}$. 
 		 In the inset we show the raw data $M_g(t)$ versus time for different system linear sizes $L$. 
		 \label{fig_mag}
		} 
\end{figure}

As explained in Sec.~\ref{conf}, the critical geometric clusters of the $2d$IM are well understood as they are the FK
clusters  of the tricritical $q=1$ Potts model. However, it is not obvious whether this correspondence should hold in an
out of equilibrium situation, typically a quench towards the critical point.  In particular, one may wonder whether the
dynamical exponent $z$ for spin clusters is the Ising one. To check this reasonable assumption, we calculated specific
quantities that allow us to extract spin clusters exponents. For example, in equilibrium, the size of the largest spin
cluster scales as $L^{-(\beta/\nu)_{tri}}$ where $(\beta/\nu)_{tri}=5/96$ and $\beta_{tri}$ is the magnetic exponent of
the tricritical $q=1$ Potts model, $\nu_{tri}$ is the exponent associated to the divergence of the correlation length
and $L$ is the linear size of the sample. We denote this quantity $M_g$. We now consider a quench from $T=0$ to $T_c$
for different system linear sizes $L$, and we compute $M_g(t)$. We consider this quench since the scaling form is simple
in this case and the dynamical exponent should not depend upon the initial condition as long as we quench at the
critical temperature. The inset of Fig.~\ref{fig_mag} shows the relaxation of this quantity for several sizes. We then
apply the following scaling $M_g(t)\rightarrow M_g(t)L^{(\beta/\nu)_{tri}}$ and $t\rightarrow t/L^z$.  The collapse of
the different curves giving access to $z$ is presented in Fig.~\ref{fig_mag}.  The value obtained converges towards the
most accurately estimated value $z=2.1667 \ (5)$ \cite{nightingale_monte_2000}. This gives strong evidence that the
dynamics of the spin clusters are indeed governed by the Ising dynamical exponent $z$.  In consequence we will use the
same $z$ for all the quantities computed in this work.

\subsubsection{Hull-enclosed area number density}

\begin{figure}[H] 
	\captionsetup[subfloat]{labelformat=empty}
	\begin{center} 
		\begin{tabular}{c} 
		\includegraphics[scale=1.4]{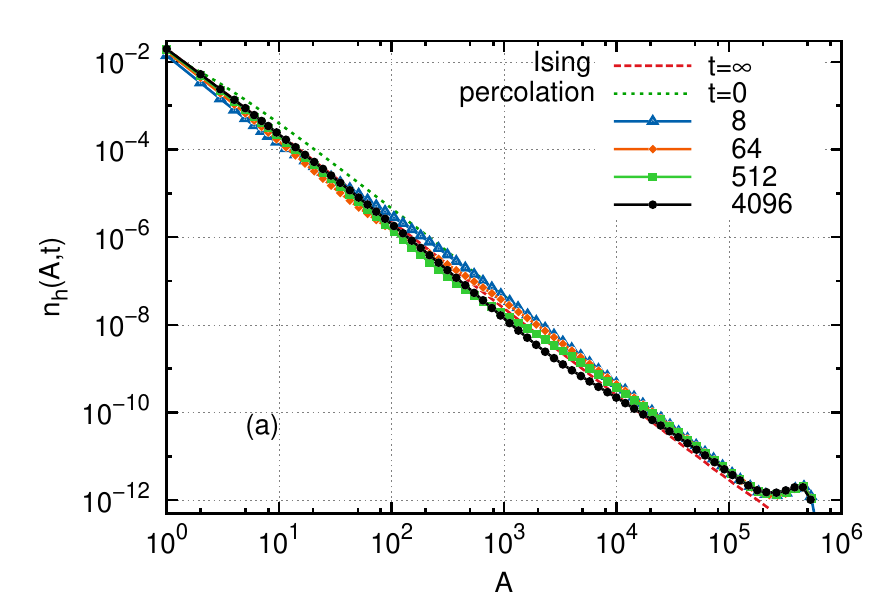}\\
		\includegraphics[scale=1.4]{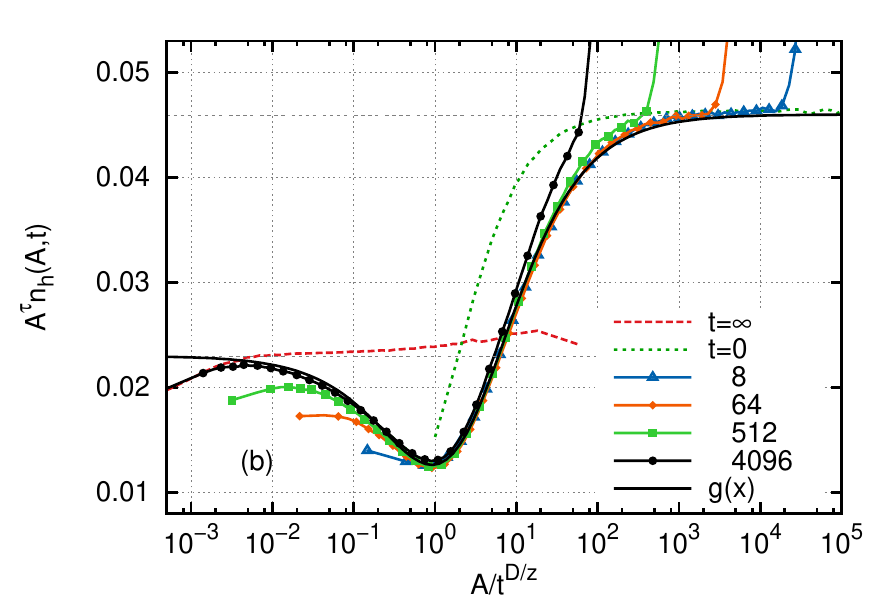}
		\end{tabular}
	\end{center} 
	\caption{Spin clusters hull-enclosed area number densities in the $2d$IM on a triangular lattice are presented.  
	(a) With dashed (red and green) lines equilibrium data at $T\to\infty$ and $T_c$. With solid lines the raw dynamic 
	data for the times given in the key (in MCs) including the spanning clusters using a double logarithmic scale. 
	(b) The areas in the x-axis are rescaled 
	by the factor $t^{\alpha}$, with $\alpha=0.92$ consistently with the expectation $\alpha=D/z=2/z$. The 
	distributions in the y-axis (in linear scale) are multiplied by $A^\tau=A^2$. The grey dotted horizontal lines at $0.046$ and 
	$0.023$ are the theoretical predictions for the $T\to\infty$ critical site percolation point and the $T_c$ one,
	respectively, and $g(x)$ is the fitting function defined in eq.~(\ref{eq:def_g}), and discussed in the text.
	\label{dist_NAA}
	} 
\end{figure}

In Fig.~\ref{dist_NAA} we present the distribution of the hull-enclosed areas, $n_h(A,t)$, at different times after the
quench.  In the first panel, Fig.~\ref{dist_NAA}(a), we display the distributions, $n_h$, versus the hull-enclosed area
$A$. As expected from the results of Cardy and Ziff~\cite{cardy_exact_2002}, the slopes of the initial  and
asymptotical (equilibrium at $T_c$) distributions  equal $-2$ in a double logarithmic plot but the prefactors are
slightly different.  For this reason, the changes occurring in the distributions are very small. Equilibrium data
(generated with a special purpose algorithm at $T_c$) are shown with (green at $T\to\infty$ and red at $T_c$) dashed
lines. The bumps at large sizes correspond to the spanning clusters with a linear size of the order of $L$, the linear
size of the system, are finite size effects. Note that the position of the bump and the very last part of the $n_h(A,t)$
close to it, are time-independent (within our numerical accuracy and for the times accessed in the simulation).

For hull-enclosed areas dynamic scaling suggests  
\begin{equation}
n_h(A,t) \simeq 
A^{-\tau} \ 
g\left( \frac{A}{t^{\alpha}} \right)
\label{eq:def_g}
\end{equation}
for $A_0 \ll A \ll L^2$ with $A_0=a^2$ a microscopic area scale and $L^2$ the macroscopic one.  To make the notation
lighter we did not add any sub-script to $\tau$ and $\alpha$.  The exponents are given by
\begin{equation}
\tau = 1+ d/D, \qquad\qquad \alpha=D/z .
\end{equation}
Since these areas are regular, i.e. they have no holes, $D=d=2$
and
\begin{equation}
\tau = 2 , \qquad\qquad \alpha = 2/z .
\end{equation}
Consistency with the asymptotic limit requires 
$g(y\to 0) = C$ and $g(y\to\infty)=2C$.

In order to test the dynamical scaling hypothesis, we impose the value $\tau=2$ and rescale the areas by a factor
$t^\alpha$ with $\alpha$ a free parameter whose value is determined by the collapse of the numerical data on a master
curve. The best collapse is obtained for $\alpha=0.92\ (5)$ and is seen in Fig.~\ref{dist_NAA}(b) with the theoretical
percolation and Ising critical distributions being just horizontal after the given ordinate rescaling. The numerical
equilibrium critical distributions (red dashed line for $T_c$ and green dashed line for $T\to\infty$) have been placed
at convenience (horizontally) to ease the comparison with the other distributions. The horizontal dotted lines
correspond to the exact prefactors $C=1/(16\sqrt{3}\pi)$ and $2C$ for the critical Ising and percolation distributions,
respectively. The value of $\alpha$ is in agreement with the expected value $\alpha=2/z\simeq0.92306$ with
$z\simeq2.1667 \ (5)$ \cite{nightingale_monte_2000} since the hull-enclosed areas scale as the square of the dynamical
length scale $\xi(t)\sim t^{1/z}$.  It is quite remarkable that, while no time-dependent rescaling has been applied to
the distribution, the curves collapse so accurately in the vertical direction. One may remark that the scaling is not so
good for times that are smaller than 64~MCs, which is not surprising since the dynamical scaling hypothesis only holds
after a non-universal time-scale. 

In curvature-driven coarsening we know from~\cite{Sicilia2007,Sicilia2009}, that $n_h(A,t)=2C/(A+\lambda_h t)^2$ if we use
an infinite temperature initial condition, i.e. $n_h(A,0)=2C/A^{2}$, with $\lambda_h$ a non-universal parameter. It
implies 
\begin{equation}
A^{\tau}n_h(A,t)=f(A/\lambda_h t) \quad \mbox{with} \quad f(x)=\frac{2C}{(1+1/x)^2}.
\end{equation}
$f(x)$ looks like the right part of $A^{\tau}n_h(A,t)$ for $A>t^{2/z}$ in our case. Since we consider a situation with
two critical points and not only one as in the sub-critical case we are tempted to think that the minimum of $g(x)$
around $x\simeq 1$ and more generally its non-monotonic behaviour can be 
\noindent reproduced with a sum of two functions similar to $f(x)$,
one decreasing from $C$ to $0$ and one increasing from $0$ to $2C$. This suggests for $g(x)$ defined in
eq.~(\ref{eq:def_g}) the form:
\begin{equation}
g(x)=C\left(\frac{1}{(1+ax^b)^c}+\frac{2}{(1+(ax^{b})^{-1})^c}\right). 
\end{equation}
We left the powers $b$ and $c$ as free parameters as there is a priori no reason for them to be simple integers as in
the sub-critical coarsening situation. Numerical inspection of the data shows a good fit for the values $a\simeq0.65$,
$b\simeq0.79$ and $c\simeq 2.3$.  Roughly speaking, $a$ fixes the position of the minimum, $b$ its width and $c$ its
depth. Note that $c$ is chosen to allow the fitting curve $g(x)$ to go through the minimum of the numerical data but we
do not attribute a special meaning to this value. Indeed the expansion of $g(x)$ close to the asymptotic values $C$ and
$2C$ is proportional to $x^b$ and $x^{-b}$, respectively, both independant of $c$.

\subsubsection{Hull-enclosed areas and hull lengths}
\begin{figure}[H] 
	\captionsetup[subfloat]{labelformat=empty}
	\begin{center} 
		\begin{tabular}{c} 
		 \includegraphics[scale=1.40]{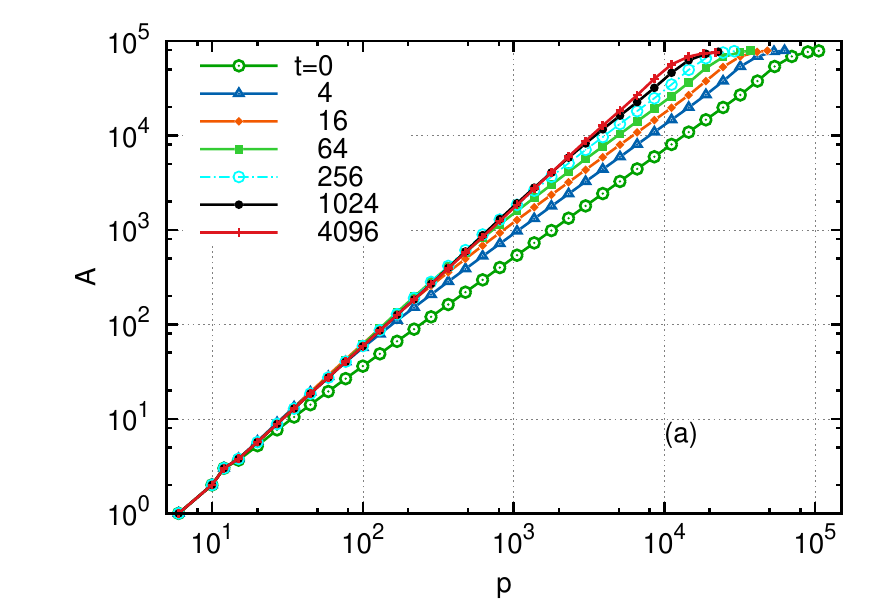}\\
		 \includegraphics[scale=1.40]{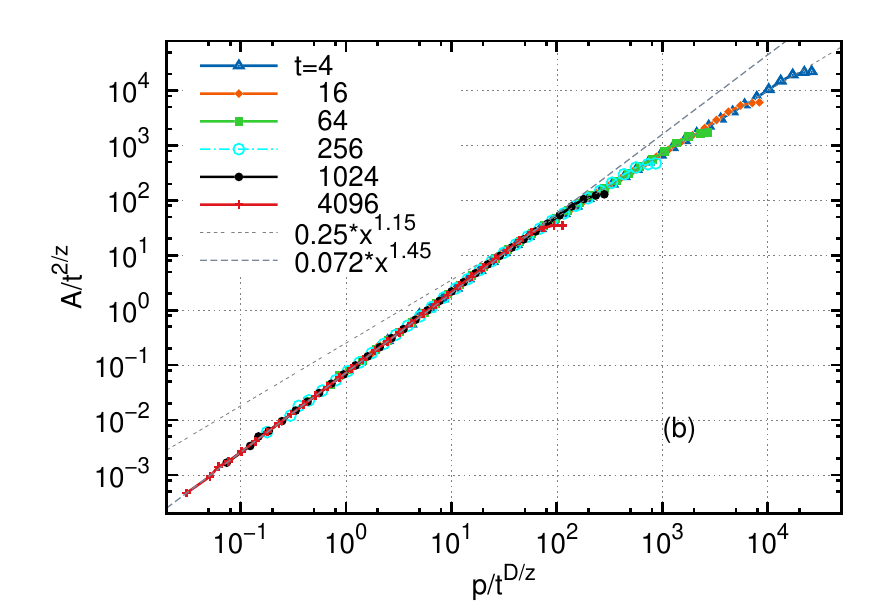}
		\end{tabular}
	\end{center} 
	\caption{The hull-enclosed area against the length of the hull at different times after the critical quench for
		 $L=400$. (a) Raw data. (b) Scaled plot. The master curve crosses over from $x^{d/D_h^{I}}$ to
		 $x^{d/D_h^{P}}$ with $d/D_h^{I} \simeq 1.45$ and $d/D_h^{P} \simeq 1.15$.
		}
	\label{AvsP}
\end{figure}

As in~\cite{Sicilia2007,Sicilia2009,loureiro_geometrical_2012} we study the relation between the hull-enclosed areas and the hull
lengths by tracing an averaged scattered plot of $A$ against $p$ in Fig.~\ref{AvsP}. In panel (a) we show the raw data
for different times after the critical quench. In panel (b) we scale the data by the relevant typical growing scales. In
the case of the hull-enclosed areas this is the linear growing length, $\xi(t)$, to the power of their fractal dimension
that is simply $D=d=2$ for these regular objects. In the case of the hull lengths, instead, the relevant linear scale is
still the growing length, $\xi(t)$, now to the power of the hulls fractal dimension in the critical Ising point, which
is $D^{I}_h=11/8$, as given in Table~\ref{tab_exp}. The master curve shows a clear cross-over between two power law
behaviors with the powers $D/D_h^{I}=16/11\simeq 1.45$ controlling the small scales and $D/D_h^{P}=8/7\simeq 1.14$
controlling the large scales. Summarizing,
\begin{eqnarray}
\displaystyle{
\left( \frac{A}{t^{D/z} } \right) \simeq \left( \frac{p}{t^{D_h^{I}/z}} \right)^{\zeta}
\qquad
\mbox{with} \qquad \zeta \ = \
\left\{
\begin{array}{l}
D/D_h^{I} \qquad \mbox{for} \qquad   p/t^{D_h^{I}/z} \ll 1 , 
\\
D/D_h^{P} \qquad \mbox{for} \qquad p/t^{D_h^{I}/z} \gg 1 .
\end{array}
\right.
}
\label{eq:Avsp}
\end{eqnarray}

\subsubsection{Hull length number density}

\begin{figure}[H] 
        \captionsetup[subfloat]{labelformat=empty}
        \begin{center} 
        	\begin{tabular}{c} 
        	\subfloat[\label{dist_NP1}]{\includegraphics[scale=1.40]{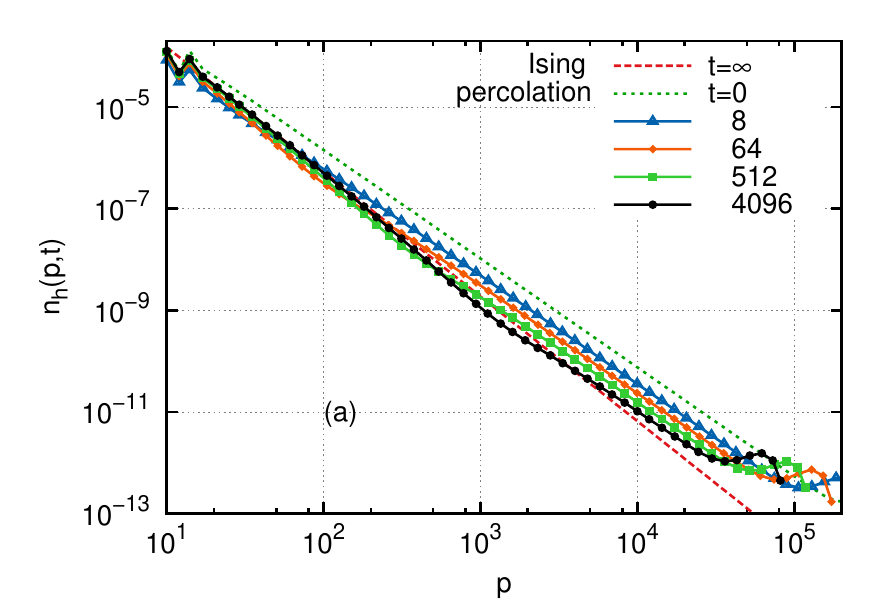}}\\
        	\subfloat[\label{dist_NP2}]{\includegraphics[scale=1.40]{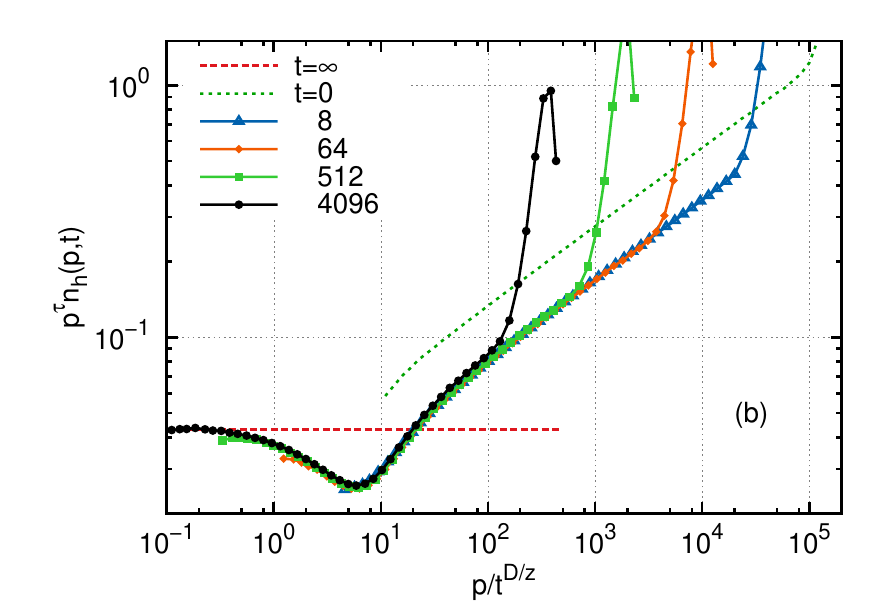}}
        	\end{tabular}
        \end{center} 
        \caption{Time-dependent distribution of the hull length $p$ of the spin clusters, 
        in a double logarithmic scale for the times given in the key in MCs. 
        (a) Raw data dynamic data together with the ones for equilibrium at the departure and arrival values, $T\to\infty$ and 
        $T_c$, that are shown with dashed (red and green) lines. (b) Same as in (a) but the distributions are multiplied by $p^\tau$ 	
        with $\tau=27/11$ the exponent of the equilibrium number density at $T_c$  and the lengths are divided by 
        $t^\alpha$ with $\alpha=D_h/z \simeq 0.63$ with $D_h$ the fractal dimension of the equilibrium hulls at $T_c$.
        \label{dist_NP}} 
\end{figure}

Next we proceed with the hull length distribution displayed in Fig.~\ref{dist_NP}.  In the first graph,
Fig.~\ref{dist_NP}(a), we show the distributions $n_h(p,t)$ vs. $p$  for various times after the quench where $p$ stands
for the hull length. For this quantity, the equilibrium behavior for the critical Ising model is $n_h(p)\sim~p^{-27/11}$
and it is different from percolation criticality where $n_h(p)\sim~p^{-15/7}$. These equilibrium curves are drawn with
dashed lines in the figure.  We see in Fig.~\ref{dist_NP}(a) that the dynamic curves interpolate between these two
critical laws.  For example, we can observe that for $t~=~4096$~MCs there is a qualitative change for $p~\sim~2000$.
However, since the difference between $27/11$ and $15/7$ is small it is still difficult to get a precise picture of what
is happening with this data representation.  Note that, contrary to what happens with the area number densities, the
time-dependence is seen in the full extent of the curves and not only for (relatively) small scales. Even the bump is
displaced towards shorter lengths in the course of time.

A better description is given in Fig.~\ref{dist_NP}(b) where we drew the same distribution multiplied by $p^\tau$ with
$\tau={27/11}$, the exponent of the tail in the equilibrium distribution at $T_c$, so that the Ising equilibrium
critical distribution is horizontal. As before, the lengths are rescaled by a factor $t^\alpha$ to take into account the
growing length scale. Searching the value of $\alpha$ that gives the best horizontal collapse we find $\alpha=0.63\
(3)$. This fits well with the value expected from dynamical scaling argument such that $p$ is rescaled by $\xi(t)^{D_h}$
with $D_h=D^{I}_h=11/8=1.375$ the fractal
\vspace{1cm}
\noindent  dimension of Ising clusters' hulls.  $D_h/z\simeq0.6348$ is in good agreement
with the measured value and this clearly supports the idea that the growing clusters have a fractal boundary. It is
interesting to notice that this does not happen in sub-critical coarsening where the domain growth is curvature-driven
and the boundaries of clusters that are smaller than the cross-over scale are smooth and do not have a fractal structure
(see~\cite{Sicilia2007} e.g. for the study of such a case in the $2d$IM). The asymptotic value, for $p/t^{D_h/z} \ll
1$, coincides with the expected equilibrium value, as obtained from the equilibrium simulations and shown with the (red)
dashed horizontal line.
 
The behavior in the large scale or short time limits, $p\gg t^{D^{I}_h/z}$, is also interesting. Since there is a
one-to-one relation between areas and perimeters, we can use $n_h(A,t) dA \simeq n_h(p,t) dp$ where $A$ and $p$ are
related by eq.~(\ref{eq:Avsp}) (note that we use the same symbol $n_h$ to represent two different functions). Using
$n_h(A,t)\simeq A^{-\tau}$ with $\tau=1+d/D$ for large values of $A$ and the value of the exponent $\zeta$ given in
eq.~(\ref{eq:Avsp}) for large scales we find
\begin{equation}
n_h(p,t) p^{\tau^{I}_h} 
\simeq 
\left(\frac{p}{t^{D_h^{I}/z}}\right)^{\gamma}
\qquad \mbox{with} \qquad
\gamma = d \ \frac{D_h^{P} - D_h^{I} }{D_h^{P} D_h^{I} }=\tau_h^I-\tau_h^P
\label{eq:gamma}
\end{equation}
independently of $D$. We used the super-script $I$ to stress that the $\tau$-value in the factor in the left-hand-side
corresponding to critical Ising equilibrium was used in the scaling of data in Fig.~\ref{dist_NP}(b). The bump at the end
of the data is just the contribution of the spanning clusters. The growing part of the  master curve is very well
described by the power law given above. Indeed, close to it we placed the data obtained in equilibrium at $T\to\infty$
(dashed green line) that in the representation used in the plot is given by $x^\gamma$, with the value of $\gamma$ given
in eq.~(\ref{eq:gamma}). The two curves are parallel in the selected range of variation (within numerical accuracy)
confirming our prediction.

\subsubsection{Domain area number density}

We have checked that the area-hull relation for the domains conforms to the scaling in eq.~(\ref{eq:Avsp}) with the relevant
$\zeta$ given by the fractal dimension of the domains.

\begin{figure}[H] 
	\begin{center} 
	\captionsetup[subfloat]{labelformat=empty}
	\begin{tabular}{c} 
		\subfloat[\label{dist_NA1}]{\includegraphics[scale=1.4]{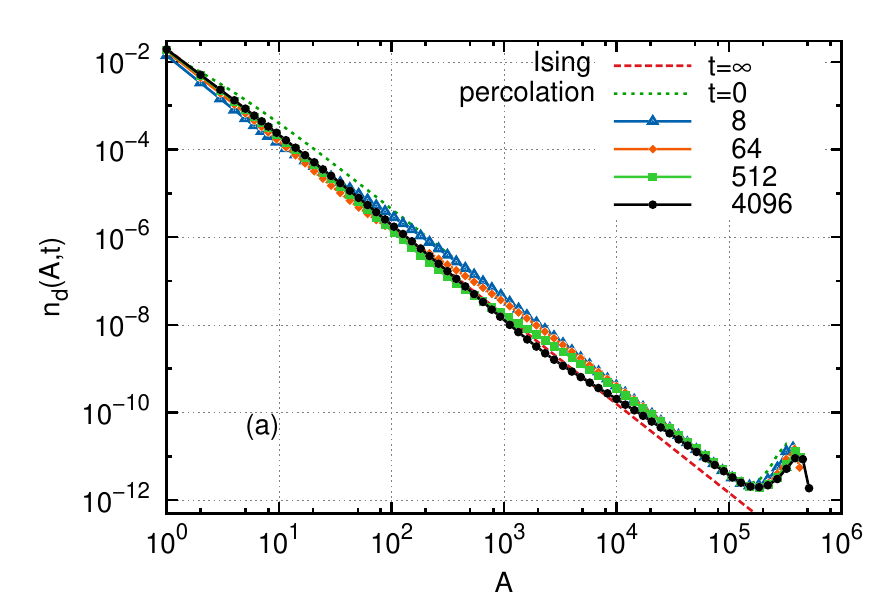}}\\
		\subfloat[\label{dist_NA2}]{\includegraphics[scale=1.4]{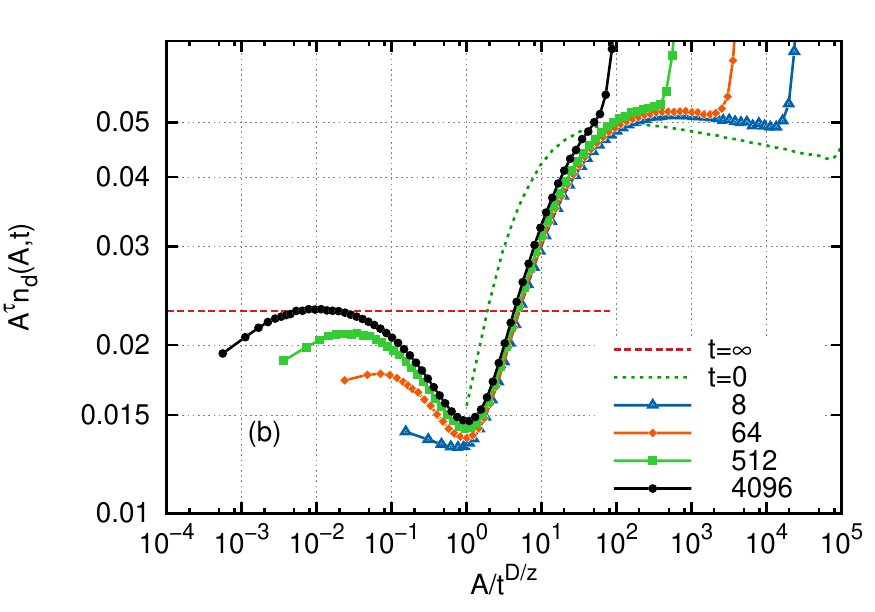}}
	\end{tabular}
	\end{center} 
	\caption{Time-dependent domain area distribution. 
	(a) Raw data at the times given in the key in MCs. (b) Same as in (a) with the distributions multiplied by
	$A^\tau=A^{2.025}$ and the area rescaled by $t^{\alpha}$ with $\alpha=D/z\simeq 0.90$.
	\label{dist_NA}} 
\end{figure}

Finally, we present the spin clusters (or domain) area distributions at several instants after the quench in
Fig.~\ref{dist_NA}.  Again the two equilibrium distributions appear as (green and red) dashed lines.  Even for the
relatively short time $t=8$~MCs the change in the distribution is rather drastic, and we can observe that the system has
been depleted from small clusters with area up to $80$ because of the onset of the interactions. This is related to the
fact that the fractal dimension of the Ising clusters at equilibrium is greater than the dimension of the percolation
clusters. Qualitatively we can explain this fact by saying that the Ising clusters are less porous because of the
interactions. We perform again the rescaling of the areas by $t^\alpha$ and we multiply the distributions by $A^{\tau}$,
with $\tau=379/187\simeq2.027$ being the critical Ising exponent for the distribution of domain areas (see the Table).
Here the best collapse (Fig.~\ref{dist_NA}(b)) is obtained for the value $\alpha=0.90\ (3)$. The fractal dimension of
the geometrical clusters at the critical point is $D_c= 187/96\simeq1.948$.  Then the value of $\alpha$ is in good
agreement with the value expected from scaling arguments $D_c/z\simeq 0.899$.  Note that if we had neglected the fractal
character and used $D_c=2$, we would have obtained $2/z\simeq 0.923$ which is also in the interval of confidence of our
numerical estimate but we think the value $D_c/z$  with $D=187/96$ is the correct one.  The scaling just discussed gives
evidence for the importance of the fractal structure of the spin clusters at the critical point. The values obtained
suggest a fractal front, whose fractality is the critical Ising one, propagating on increasingly larger scales and
bringing the system towards a new equilibrium, although very slowly, with a law governed by the dynamic exponent $z$. 

\subsection{Fortuin-Kasteleyn clusters}
\label{sec:fk_clusters}

In this Section we present the results of simulations in which we analyze the FK clusters. These clusters are critical
at $T=T_c$ but they are not critical at $T \to \infty $ since at infinite temperature $p=1-e^{-2\beta}=0$ and they are
all composed of only one site. We constructed the dynamic FK clusters adapting the usual procedure explained in
Sec.~\ref{subsec:definitions}  and sketched in Fig.~\ref{def:fk} to the time-dependent domains. At different instants
after the quench from $\beta=0$ to $\beta_c$ we used the probability $p=1-e^{-2\beta_c}$ to erase bonds from the spin
domains existing at the chosen times.  We next studied various distribution functions associated to the FK clusters.  We
recall that in this study we used a square lattice for simplicity and that the initial point (infinite temperature) is
\emph{not} critical for the spin clusters either.

We have checked that the number density of hull-enclosed and domain FK areas as well as the area-perimeter relations satisfy the
expected scaling relations already discussed for spin clusters in Sec.~\ref{sec:spin_clusters}, so we prefer to show the
data for other observables.

\subsubsection{Hull length}

In Fig.~\ref{fk:npfk}(a) we show the hull length distributions at $T\to\infty$ and at $T_c$ as well as the
distributions of the same quantities at several times after the quench. We can easily see that the initial distribution
is not a critical one since it is not a straight line in this log-log graph. As time elapses, the distribution
approaches a power law behavior since it heads towards criticality. Again, to get a better description, the curves are
collapsed on Fig.~\ref{fk:npfk}(b) by using the now usual rescaling $p^{\delta} n_h(p,t)= f(p/t^\alpha)$. The best
collapse onto a master curve is obtained for $\delta=2.12$ and $\alpha=0.87$.  These values are compatible with $\tau=2.2$
and $D_h/z\simeq 0.77$, the values at the Ising critical point, cf. Table~\ref{tab_exp}, though here the agreement is
not as good as for the hulls of the spin clusters.  We  notice that for times longer than $2000$ MCs the curves do not
collapse anymore for large sizes. This is expected since it roughly corresponds to the moment when the dynamical
length scale reaches the order of the linear size of the system $\xi(t)\sim L$. After this time we are no longer in the
regime of unconstrained domain growth as the clusters are sensitive to the finite size of the system.

\subsubsection{External perimeter}

The same scaling is finally applied to the external perimeter distributions of the FK clusters, the idea being to check
whether the consequences of Duplantier's duality hold out of equilibrium. The curves are not shown since they are very
similar to the ones in Fig.~\ref{fk:npfk}. The curves collapse for $\delta=2.56$ and $\alpha=0.63$. To compare this
values with the related equilibrium exponent at $T_c$ we must remember that, according to the duality relation evoked
before [$D_{ep}(\kappa)=D_h(16/\kappa)$ for $\kappa \ge 4$], the external perimeters of the FK clusters at $T_c$
should scale at equilibrium like the hulls of the spin clusters, that is possess also at $T_c$ a fractal dimension
$D_{ep}(\kappa=16/3)=D_h(\kappa=3)=11/8$. This gives $\tau\simeq2.45$ which is compatible with $\delta$ and
$\alpha=D_{ep}/z\simeq 0.63$. Therefore the relation between the statistical and geometric properties of the external
perimeters of FK clusters and the hulls of spin clusters proven in equilibrium, seems to remain valid out of equilibrium
for the growing FK clusters.

\subsection{The interface}
\label{sec:interface}

\begin{figure}[ht!] 
        \begin{center} 
        \includegraphics[scale=1]{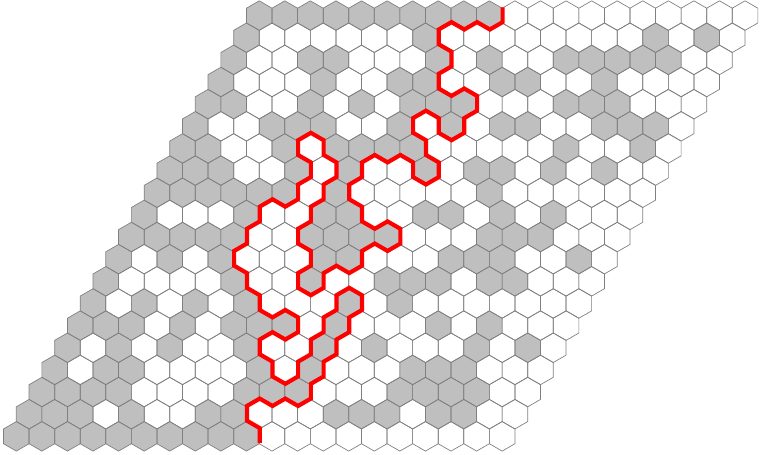}
        \end{center} 
        \caption{An interface in a $2d$IM on a triangular lattice created by 
        	imposing spin up boundary condition on the left (grey hexagons) and spin down on the right
         	(white hexagons). The curve is defined on the edges of the dual lattice, the honeycomb lattice represented in the 
	figure.\label{sle:drawing}} 
\end{figure}

\begin{figure}[H] 
	\captionsetup[subfloat]{labelformat=empty}
	\begin{center} 
		\begin{tabular}{c} 
		\subfloat[ \label{fk:npfk_no_resc}]{\includegraphics[scale=1.4]{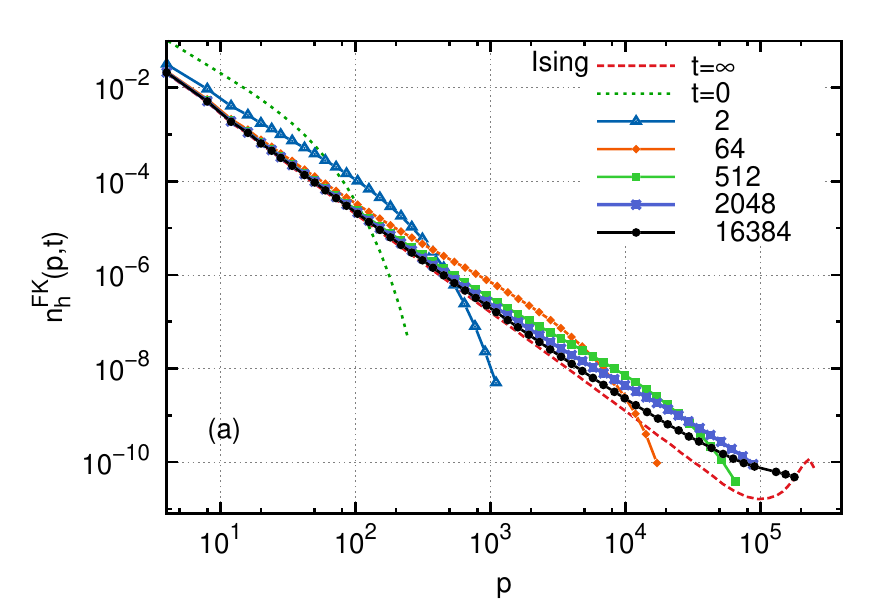}}\\
		\subfloat[ \label{fk:npfk_resc}]{\includegraphics[scale=1.4]{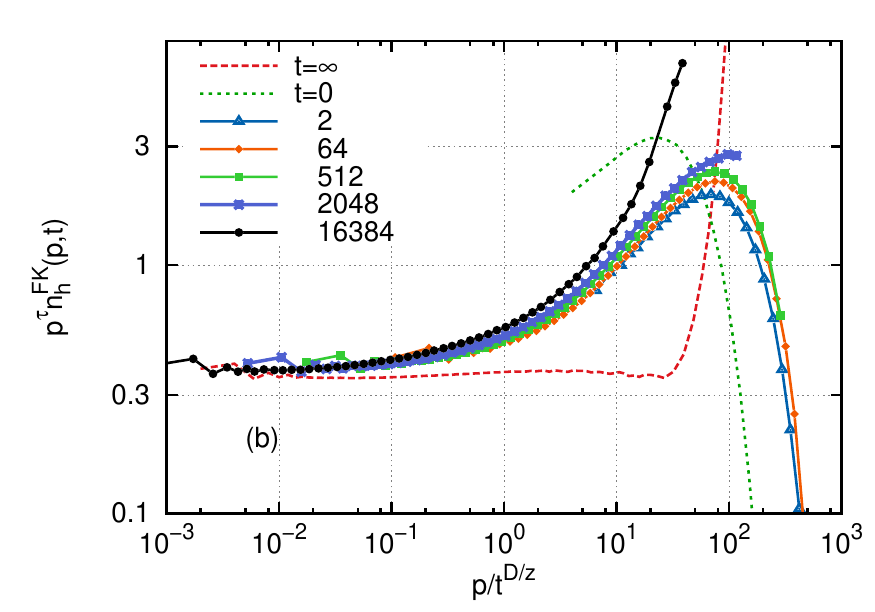}}
		\end{tabular}
	\end{center} 
	\caption{Number density of FK hull length $p$ at several times after the quench given in the key for a system of
		 linear size $L=320$. (a)  Raw data. (b) Same data presented in the rescaled form
		 $p^{\tau}n^{FK}_h(p,t)$ vs. $p/t^{\alpha}$ with $\tau=2.12$ and $\alpha=0.87$ 
		 that gives the best data collapse. \label{fk:npfk}} 
\end{figure}

In the previous sections, we have seen fractal dimensions appear while considering the dynamic scaling of several
probability distributions.  A different strategy, that has proven to be very useful in the analysis of the equilibrium
critical points and in particular in the context of the study
\noindent of SLE~\cite{Cardy05,gruzberg_stochastic_2006,Bernard06}, 
is to study the fractal properties of an artificially generated interface. This is the line of research
we follow here by generating the interface as follows.  We take a $2d$IM on a triangular lattice and we quench the
system from equilibrium at $T\to\infty$ to  $T_c$ at $t=0$. With appropriate boundary conditions we force the existence
of a unique curve, defined on the edges of the dual honeycomb lattice, separating spin clusters of opposite sign and
going, say,  from top to bottom. One such curve is shown in Fig.~\ref{sle:drawing}, their properties are analyzed in
this section.

\subsubsection{Fractal dimension}

\begin{figure}[t!,scale=0.5] 
	\begin{center} 
	\captionsetup[subfloat]{labelformat=empty}
		\subfloat[]{\includegraphics[scale=1.4]{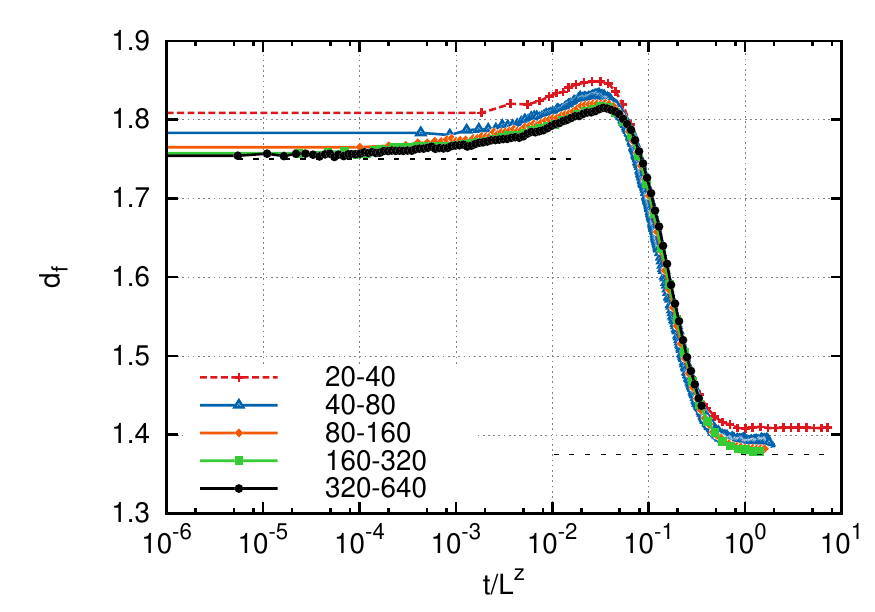}}
	\end{center} 
	\caption{Fractal dimension of an interface vs. $t/L^z$, as deduced from eq.~(\ref{eq:fractal-dim}) 
	for several couples of system sizes $L$ and $L'=2L$ given in the key. \label{sle:fractal}} 
\end{figure}

We are interested in the fractal dimension of the interface. To compute this quantity we measured the length of the
curve $l(L,t)$ with $L$ the linear size of the system  as a function of the time $t$. If the curve is fractal at time
$t$, then the fractal dimension $d_f(t)$ is defined by $l(L,t)\propto L^{d_f(t)}$. An effective $d_f(t)$ is obtained
from the slope of $l(L,t)$ between two sizes $L$ and $L'$:
\begin{equation}
d_f(L,L',t)=\frac{\ln l(L,t) - \ln l(L',t) }{\ln L-\ln L'}.
\label{eq:fractal-dim}
\end{equation}
This yields a fractal dimension depending on time but also on the two lengths $L$ and $L'$ because of finite-size
corrections. In Fig.~\ref{sle:fractal} we draw this fractal dimension vs. time $t$ rescaled by $L^z$ for several values
of $L$ and $L'$ with $z$ the dynamical exponent employed in Sec.~\ref{sec:spin_clusters} and Sec.~\ref{sec:fk_clusters}.

We observe that for short times, i.e. $t\ll L^z$, the fractal dimension is compatible with $7/4=1.75$ the fractal
dimension of the hulls at the percolation threshold, while for long times it is closer to $11/8=1.375$ the fractal
dimension of the Ising clusters' hull at the critical point. The compatibility is strong since even the finite-size
corrections for the short (long) scales correspond to the related equilibrium values. The points completely on the left
correspond to the initial values being rejected to $-\infty$ in the log scale. The dynamical scaling is also very
accurate here since the curves perfectly collapse apart from finite-size corrections.

The main disadvantage of calculating a fractal dimension this way is having to handle the two lengths in the
interpretation of the fractal dimension. This  is uneasy. We use a different and more performant 
approach in the next subsection.

\subsubsection{The winding angles}

The percolation and Ising model hulls, at equilibrium and in the continuum limit, are conformally invariant curves
described by a stochastic Loewner evolution SLE$_\kappa$ with $\kappa=6$ for percolation~\cite{smirnov_critical_2001}
while $\kappa=3$ for Ising hulls~\cite{smirnov_towards_2007,chelkak_universality_2009}. The parameter $\kappa$ can be
determined numerically by computing the variance of the winding angle for two points chosen at random on a conformally
invariant curve at a distance $l$ along the curve which equals
\begin{equation}
\langle \theta^2 \rangle=\mbox{ct} +\frac{4(D-1)}{D}\ln l ,
\label{sle:var1}
\end{equation}
where $D$ is the fractal dimension of the
curve~\cite{duplantier_winding-angle_1988,duplantier_harmonic_2002,wieland_winding_2003} and $\mbox{ct}$ is a constant.
Finally, using eqs.~(\ref{sle:dim}) and (\ref{sle:var1}), we obtain the following form for the variance of the winding angle:
\begin{equation}
\langle \theta^2 \rangle=\mbox{ct} +\frac{4\kappa}{8+\kappa}\ln l. 
\label{sle:var3}
\end{equation}

\begin{figure}[H] 
	\captionsetup[subfloat]{labelformat=empty}
	\begin{center}
		\begin{tabular}{c} 
		\subfloat[ \label{sle:quench1}]{\includegraphics[scale=1.4]{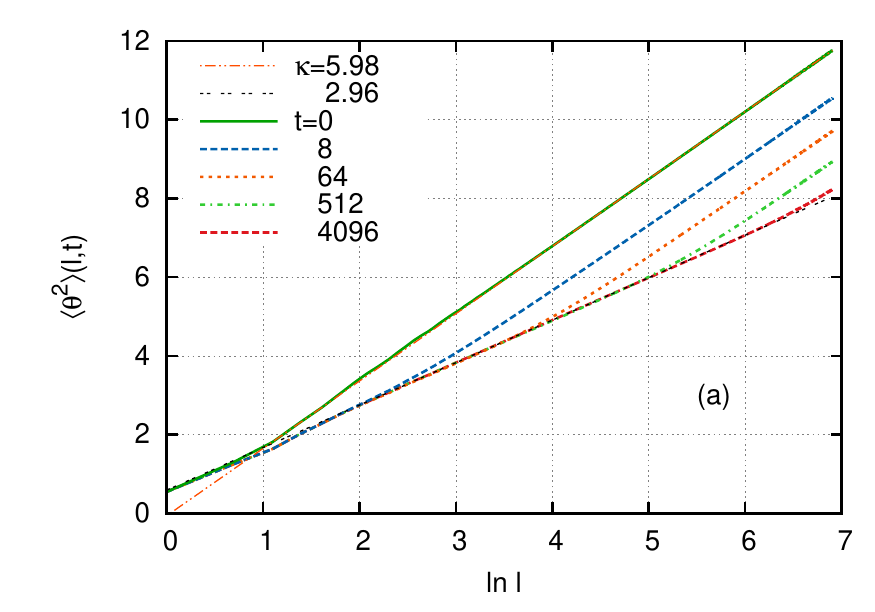}}\\
		\subfloat[ \label{sle:eq}]{\includegraphics[scale=1.4]{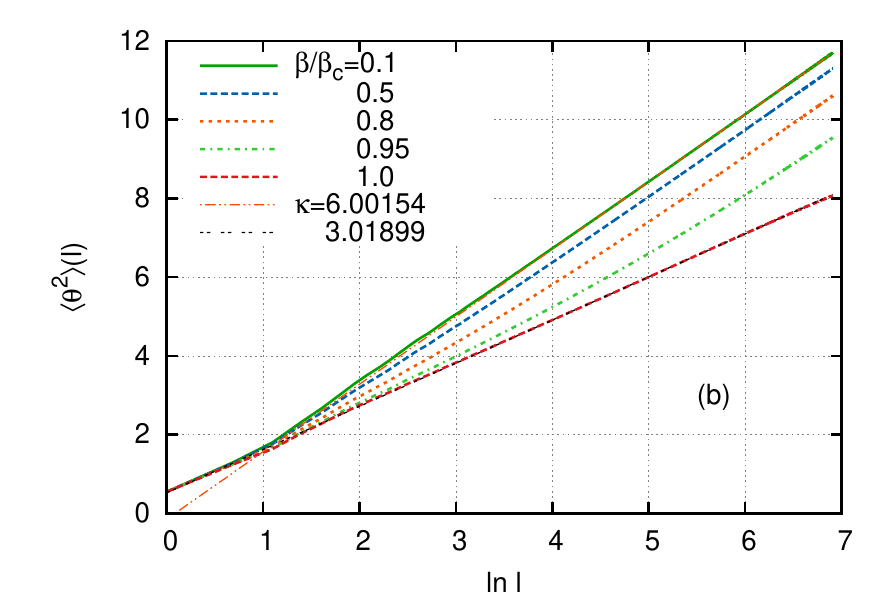}}
		\end{tabular}
	\end{center} 
	\caption{Variance of winding angle of interfaces out of equilibrium (a) and in equilibrium (b) for a system of linear size
		$L=1280$. (a) $\langle\theta^2 \rangle$ vs. $\ln l$ for different times after the quench from $T=\infty$ to $T=T_c$. 
		(b) $\langle \theta^2 \rangle$ vs. $\ln l$ at equilibrium for different inverse temperatures $\beta=1/T$ in unit of 
		$\beta_c$ given in the key.
		\label{sle:fig1} 
	} 
\end{figure}

The winding angle variance, $\langle \theta^2 \rangle (l,t)$,  is measured as a function of the distance $l$ for
different times after the quench. The results for a system of linear size $L=1280$ are given in Fig.~\ref{sle:fig1}(a).
For the time $t=0$ corresponding to percolation, a fit of the form~(\ref{sle:var3}) yields $\kappa=5.98$ in excellent
agreement with the exact result $\kappa=6$. For the largest time simulated $t=4096$~MCs, a fit of the
form~(\ref{sle:var3}) yields $\kappa=2.96$. For this fit we kept the data with $\ln l<6.5$ since for larger distances,
the variance starts deviating from a straight line. This value is in excellent agreement with the exact result for Ising
model spins clusters in equilibrium, $\kappa=3$. The best fits for $t=0$~MCs 
\begin{figure}[t!,scale=0.5] 
	\captionsetup[subfloat]{labelformat=empty}
	\begin{center}
		\subfloat[]{\includegraphics[scale=1.4]{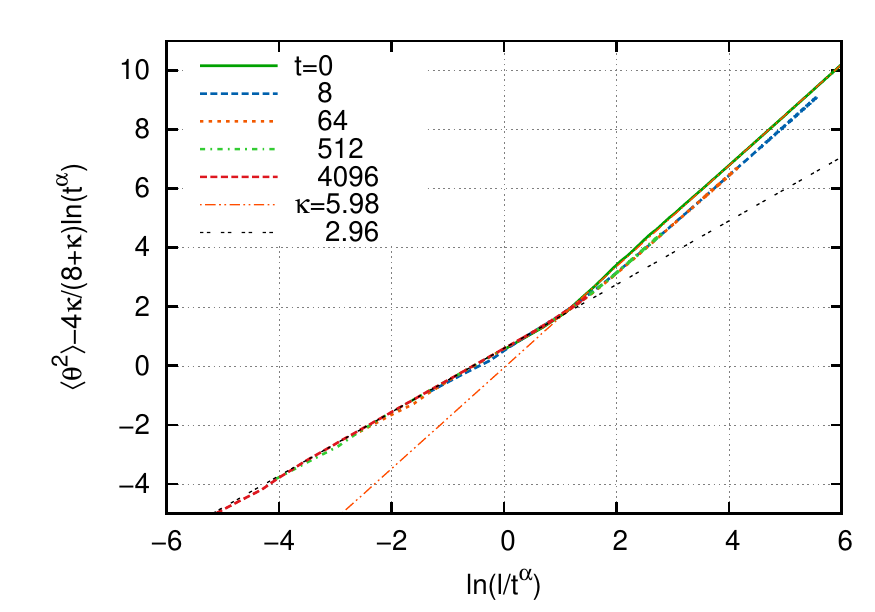}}
	\end{center} 
	\caption{Same curves as in Fig.~\ref{sle:fig1}(a) with the rescaling $l\rightarrow l/t^\alpha$ and 
		 $\langle \theta^2 \rangle(l,t)~\rightarrow~\langle \theta^2 \rangle(l,t)~-~4\kappa/(8+\kappa)\ln t^\alpha$ with $\alpha=0.65$. 
		 The dashed lines are our best fits to the small and large scale behavior and yield the $\kappa$ values given in the key. 
	         \label{sle:quench2}} 
\end{figure}

\noindent and $t=4096$~MCs are represented as dashed
lines in Fig.~\ref{sle:fig1}(a).  We now wish to collapse these curves to make the dependence of the growing length
scale upon time explicit. The length $l$ along the curve is divided by $t^\alpha$ and the winding angle variance
corresponding to a length $t^\alpha$ is subtracted. Thus, in Fig.~\ref{sle:quench2}, we plot $\langle \theta^2 \rangle
(l,t) - 4\kappa/(8+\kappa)\ln t^\alpha$ vs. $\ln(t/t^\alpha)$. The curves collapse very well for $\alpha=0.65$ which is
in good agreement with $D_h/z\simeq0.63$ with $D_h=11/8$ the fractal dimension of Ising clusters' hulls. This means
that the same rescaling as the one used for the distributions works for the winding angle variance as well. In the
latter case, however, the transition between the two equilibrium regimes is more rapid. Indeed, for distances smaller
than the characteristic length $\xi(t)^{D_h}\sim t^{D_h/z}$ the interface is, at least for the observable $\langle \theta^2
\rangle$, as if the system were in equilibrium at $T_c$ and for larger distances as if the system were still in the
independent percolation regime. In the previous section, we have observed  a similar crossover in the behavior of the
probability distributions, with small and large scales being the ones of the initial regime of percolation and the ones
of the state towards which the system is converging, respectively, but the crossover was far less abrupt. 

For comparison, we present the plot of $\langle \theta^2 \rangle$ against $\ln l$ in \emph{equilibrium} for various
temperatures in Fig.~\ref{sle:fig1}(b). It is known that the high temperature phase of the Ising model at zero field on
the triangular lattice is peculiar in the sense that it is critical and in the universality class of uncorrelated
percolation~\cite{klein_renormalization-group_1978,balint_high_2010}. This is not true for other planar lattices.  As we
have already seen, the contours of percolation clusters on the triangular lattice are described by SLE$_6$. This explains
why  for sufficiently large sizes (larger than the correlation length induced by the interactions between spins) we
recover a dependence of $\langle \theta^2 \rangle$ with $\ln l$ that is compatible with $\kappa=6$, i.e. the behavior of the
interface is governed by the attractive infinite temperature point.  

Both Figs.~\ref{sle:fig1}(a) and \ref{sle:fig1}(b) are compatible with $\kappa=6$ at large scale. However, for shorter
distances the behavior of the quenched and equilibrium curves are very different. In the case of the quench, for
sufficiently small size and up to $l~\sim~t^{D_h/z}$ the curves are compatible with $\kappa=3$ (Ising spin clusters) but
for the curves at equilibrium the critical Ising behavior is not observed at all. 

Thus during the coarsening process the interface has a radically different behaviour
from the one obtained by considering data for equilibrium at indermediate temperatures. In the first case, we flow
from $T=\infty$ to $T_c$, with an increasing scaling length $\xi(t)\sim t^{1/z}$. In the second we flow from $T_c$ to
$T=\infty$ with a correlation length $\xi~\sim~|T-T_c|^{-\nu}$ with $\nu=1$. In this second case we can only observe the
critical large scale behaviour dominated by the percolation point while in the first case we distinguish two behaviors
with a crossover controlled by the characteristic length $\xi(t)~\sim~t^{1/z}$.

\section{Conclusion}
\label{sec:conclusions}

In this paper we analyzed the dynamics of a system instantaneously quenched from one critical point to another one.
We used the prototype statistical  model, the $2d$ Ising model, quenched from $T\to\infty$, i.e. critical site percolation, 
to the Ising critical point. 

As observed in~\cite{Arenzon2007}-\cite{loureiro_geometrical_2012}  for sub-critical quenches, the typical growing length, $\xi(t)
\simeq t^{1/z}$ in this case, separates two length scales. In the smaller one, all mesoscopic observables (areas,
perimeters, etc.) satisfy the statistical and geometric properties of the working temperature equilibrium state.
Instead, in the largest scale, these same observables are the ones of the initial state. In sub-critical quenches, this
result could be shown analytically for the hull-enclosed area distribution~\cite{Arenzon2007,Sicilia2009} and it was
confirmed numerically for different quantities and in different systems.

The attraction of quenches between critical points is that, in the departing and arrival equilibrium conditions, very
powerful analytic methods (Coulomb gas, conformal field theory, SLE) allowed one to compute a myriad of geometric
properties including fractal dimensions and related critical exponents. Could these methods be extended to deduce,
analytically, the results presented in this manuscript for a particular case, the $2d$ Ising model, that we conjecture
apply to all critical quenches, at least in bidimensional systems? This is definitely an interesting question on which
we plan to work in the future.

\vspace{2cm}

\noindent{Acknowledgements \hspace{0.5cm}
LFC wishes to thank to J. J. Arenzon, A. J. Bray, M. P. Loureiro, Y.~Sarrazin and A. Sicilia for our earlier
collaboration on similar problems. We also wish to thank J. Cardy and R. Santachiara for useful discussions on the
contents of this manuscript.}

\newpage

\bibliography{amsocdbcp.bib}

\end{document}